\documentclass[11pt,a4paper]{article}
\usepackage{amsfonts,amssymb,amsmath}
\usepackage{jheppub}
\usepackage{hyperref}
\usepackage{slashed}
\usepackage{color}
\usepackage{pdfsync}
\usepackage{accents}
\usepackage{yfonts}   
\usepackage{cancel}
\usepackage{soul}
\usepackage{verbatim}

\newcommand{\R}{\mathbb R}
\newcommand{\Z}{\mathbb Z}

\newcommand\wt[1]{\widetilde #1}

\newcommand{\qu}{\overline}

\newcommand{\be}{\begin{equation}}
\newcommand{\ee}{\end{equation}}
\definecolor{purple}{rgb}{0.6,0,0.6}
\definecolor{bl}{rgb}{0.2,0.1,0.9}
\definecolor{tur}{rgb}{0.1,0.9,0.9}
\definecolor{gr}{rgb}{0.2,0.7,0.2}
\definecolor{pink}{rgb}{1,0,0.8}

\newcounter{save}
\newcounter{rmk}

\numberwithin{equation}{section}
\title{Not Doomed to Fail\note{DCPT-17/27}}

\author[a]{Anne Taormina}
\author[b]{and Katrin Wendland}

\affiliation[a]{Centre for Particle Theory, Department of Mathematical Sciences,
University of Durham\\ Lower Mountjoy, Stockton Road, Durham DH1 3LE, U.K.}
\affiliation[b]{Mathematics Institute, University of Freiburg\\ Ernst-Zermelo-Stra\ss e 1,
D-79104 Freiburg, Germany}

\emailAdd{anne.taormina@durham.ac.uk}
\emailAdd{katrin.wendland@math.uni-freiburg.de}

\abstract{In their recent manuscript ``An Uplifting Discussion of T-Duality'', 
1707.08888,
J.~Harvey and G.~Moore have 
reevaluated a mod two condition appearing in asymmetric orbifold constructions
as an
obstruction to the description of  certain
symmetries of toroidal conformal field theories
by means of automorphisms of the underlying charge lattice. 
The relevant ``doomed to fail'' condition
determines whether or not such a lattice automorphism $g$ may lift
to a symmetry in the corresponding toroidal conformal field theory without introducing extra phases. 
If doomed to fail, then in some cases,
the lift of $g$  must have double the order of $g$.
It is an interesting question,
whether or not  ``geometric" symmetries 
are affected by these findings. In the present note, we answer this question in the negative,
by means of elementary linear algebra: ``geometric"  symmetries of toroidal conformal
field theories are not doomed to fail. 
Consequently, and in particular, the symmetry groups involved in symmetry surfing the moduli space
of K3 theories do not differ from their lifts.}


\arxivnumber{1708.01563}

\notoc

\begin{document}

\maketitle

\section*{Introduction}
Symmetries are a driving force in many areas of 
mathematics and theoretical physics. They are, in particular, central in
the investigation of the recent phenomenon of 
Mathieu Moonshine and its relations to K3 theories
\cite{eot10,ch10,ghv10a,ghv10b,eghi11,tawe11,ga12,tawe12,tawe13,so17,we17}.
The Torelli theorems for complex tori \cite{lope81} and
K3 surfaces \cite{ku77,lo81,na83,si81,to80} 
allow a description of the symmetries of these complex surfaces
in terms of automorphisms of the respective lattices of integral cohomology. 
Inspired by these theorems,
the use of lattice automorphisms 
has been extended
to the investigation and classification of symmetries 
of  superconformal field theories whose targets are  
complex surfaces \cite{ghv12,vo14}.

At the basis of this  approach lies, roughly speaking, 
the description of the moduli
spaces of K3 theories \cite{se88,ce91,asmo94,nawe00}, 
on the one hand, and of toroidal conformal field theories \cite{cent85,na86},
on the other, as Grassmannians that are modelled on the even or odd part
of the total  cohomology of the complex two-dimensional
target manifold. 
This may be viewed as an extension of the description of the moduli
spaces of hyperk\"ahler structures for the respective complex surfaces
in terms of Grassmannians, which are modelled on the second 
cohomology. 
It is thus natural to expect the properties of automorphisms of the
underlying lattices of integral cohomology 
to allow for a generalization to the conformal field theory setting.
Though well rooted in mathematics, the very description of general
symmetries both of toroidal and of K3 theories in terms of 
such lattice automorphisms  is not immediate. The more 
important are the recent results \cite{hamo17}
by J.~Harvey and G.~Moore,  which show that the symmetries of toroidal
conformal field theories cannot, in general, be fully described in
terms of their action on the underlying charge lattice. 
Applied to toroidal $N=(4,4)$ superconformal field theories at 
central charges $c=\qu c=6$, where the charge lattice may be
directly related to the lattice of odd integral cohomology of the target
\cite{di99,nawe00}, this
raises some interesting questions about the
traditional discussions of symmetries for these models.

More precisely, this discussion concerns symmetries of a toroidal conformal field
theory which induce an action on the underlying charge 
lattice fixing the respective parameter point in the Grassmannian description
of the moduli space. In \cite{hamo17}, it is shown that vice versa, for certain
toroidal conformal field theories, there exist 
automorphisms of the charge lattice which preserve 
the parameter point in the moduli
space, but which can only lift to symmetries 
of at least
double the order for the corresponding
conformal field theory. In other words,
the charge lattice of such a theory does not fully capture
the symmetry group of the conformal field theory.
There are also cases where a lift of the same order exists, 
which however  acts non-trivially on some winding-momentum fields
associated to invariant
 charge vectors\footnote{We thank G. Moore for 
emphasizing this point to us.}. Both phenomena could potentially cast some doubt
on traditional descriptions of symmetries in conformal field theory.
Whether or not  either of these phenomena occurs for a given lattice automorphism
is encoded by equation (2.17) of the manuscript
\cite{hamo17}, simply dubbed the ``doomed to fail'' condition. 
This condition had previously been stated with different 
interpretation by K.S.~Narain, M.H.~Sarmadi and C.~Vafa
\cite{nsv87,nsv91}, and in a slightly different context by J.~Lepowsky
\cite{le85,bhl06}. Hai Siong Tan used a similar approach in \cite{ta15}
to determine consistent asymmetric orbifold group actions.
The condition can be traced back to the
properties of involutions on the charge lattice which
fix the parameter point:
whenever there exists a charge vector
which  has an odd scalar
product with its image under such an involution, the 
lift is ``doomed to fail'', i.e.\ either only a  lift 
of order four exists, or there are invariant charge vectors whose associated
momentum-winding fields are multiplied by $(-1)$. 
The latter phenomenon 
of ``non-trivial phases'' is compatible with the description
of symmetries of toroidal sigma models \cite{vo14}, and it had been taken into
account in previous discussions of symmetries in conformal field theory
\cite{ghv12,vo14,gtvw14}. The former phenomenon, which affects the relevant
groups of symmetries,
has not been discussed in this form before.
The criterion for at least one of these two phenomena to occur
is very elegant, as it makes it 
a simple matter of linear algebra to check whether
or not a given lattice automorphism is doomed.

These findings could well be of direct relevance for the discussion of 
Mathieu Moonshine, independently of the general doubts that they
may cast on the use of lattice techniques in describing symmetries
of conformal field theories. Indeed,
by a $\Z_2$-orbifolding, any toroidal $N=(4,4)$ superconformal field theory
at central charges $c=\qu c=6$ gives rise to a K3 theory.
The $\Z_2$-orbifolding procedure thereby induces a map
between the respective moduli spaces which was determined  in \cite{nawe00}.
This map is described in terms of the underlying lattices of integral cohomology,
and it
requires a transition between the odd and the even integral cohomology
of complex two-tori by means of triality 
\cite{nawe00}.\footnote{Note that at this point, one needs to work on a $2\colon1$
cover of the moduli space of toroidal superconformal field theories
in order to keep the target space orientation, as is required for the
resulting K3 theories \cite[(1.17)]{nawe00}. }  The symmetries of the
underlying toroidal conformal field theories thus induce symmetries of the
resulting K3 theories. A failure of a lattice automorphism to fully capture
a symmetry of a toroidal theory may descend to the corresponding K3 theory.

We are not able to predict the full scope of consequences 
for Mathieu Moonshine that may follow from \cite{hamo17}. However, in this note
we come back  to a remark made  by J.~Harvey 
and G.~Moore in the first version of their recent paper, addressing
\textsl{symmetry surfing}.
The latter is a technique that we have proposed first in \cite{tawe10}
and that
allows to combine \textsl{geometric symmetry groups} from distinct points in the 
moduli space of K3 theories \cite{tawe11}. In \cite{tawe13},
we have shown that the maximal subgroup
$\Z_2^4\colon A_8$ of $M_{24}$ is 
the group which, by means of symmetry surfing, 
combines all geometric symmetries
induced by the  symmetries of complex tori in the corresponding
$\Z_2$-orbifold conformal field theories. By constructing the leading order
massive representation which 
contributes to Mathieu Moonshine, for this maximal subgroup,
our work \cite{tawe12} provides the first piece of
evidence that the relevant representations of Mathieu Moonshine 
might arise intrinsically from conformal field theory. Our arguments have,
in the meantime, been vastly generalized by M.~Gaberdiel, Ch.~Keller 
and H.~Paul \cite{gakepa16}, yielding additional evidence in favour of the proposal
of  symmetry surfing.

Whether or not the  symmetries that are relevant for the works 
\cite{tawe11,tawe13,tawe12,gakepa16} are doomed to fail
is an important question  for our programme. 
In this note we show that thanks to the very 
elegant ``doomed to fail'' condition, this question may be 
answered in the negative by means of elementary linear 
algebra: symmetry surfing is \textsl{not} doomed to fail by the uplifting
properties of lattice automorphisms. It is important to appreciate that
this statement is closely tied to the fact that symmetry surfing only
proposes to combine \textsl{geometric symmetries} of K3 theories. In fact,
we show more generally that \textsl{geometric symmetries} of toroidal conformal
field theories are never doomed to fail. The very definition of \textsl{geometric symmetries} needs to
be treated with great care. It refers to symplectic automorphisms of 
finite order that also leave invariant the B-field, viewed as a real-valued
two-form. In particular, this notion excludes the identification of a
B-field with its shifts by integral cohomology classes, thus excluding some
symmetries that are certainly in the realm of geometry\footnote{A lower-dimensional toroidal 
example is a  reflection in a simple root 
in the $\mathfrak{s}\mathfrak{u}(3)$-point
for two free bosons, discussed in
\cite[\S 4.1]{hamo17}.
We thank G.~Moore
for this comment.}. The adjective \textsl{geometric}
thus solely means that such 
symmetries leave invariant a geometric interpretation. 
In the context of symmetry
surfing, our notion of \textsl{geometric symmetries} is
inseparably connected with the idea that there is a 
space of states that generically exist in all K3 theories, which
bears all the structure that is relevant to Mathieu Moonshine.
As already predicted in \cite{tawe13,we14}, the cohomology of 
the chiral de Rham complex of \cite{msv98,bo01,boli00,goma03,lili07,bhs08} is 
expected to model this ``space of generic states''. 
In identifying the cohomology of the chiral de Rham complex with the large
volume limit of a topological half-twist of K3 theories \cite{ka05}, we need to 
require compatibility of the symmetries in question with a large volume limit.
We do so by requiring that they leave invariant a geometric interpretation 
of the model.
Recently, the idea  of explaining Mathieu Moonshine by means of such a space
of generic states
has been further substantiated by the observation that the cohomology
of the chiral de Rham complex for K3 surfaces indeed decomposes into
irreducible representations of the 
``small'' $N=4$ superconformal algebra at central
charge $c=6$ with proper multiplicity spaces of every massive
representation, i.e. without the occurrence of virtual representations
\cite{so17,we17}. 

We  emphasize that symmetry surfing has not been
proved to explain Mathieu Moonshine, so far, and that this proposal
may still ultimately fail, as is extensively discussed in 
\cite[\S4.5]{we17}. However,  in  this note we  use plain scientific arguments to prove
that the symmetry surfing programme is not jeopardized by misidentification of
geometric symmetries at the  level 
of toroidal conformal field theories.
Nevertheless, we expect that the findings of \cite{hamo17}
have important implications on the discussion of symmetries of K3
theories, and thus on the broader programme.

The remainder of this note is divided into two sections. 
Section \ref{geosymm} is devoted to a more detailed 
discussion of our notion of \textsl{geometric symmetries}. 
We describe the properties of the symmetries
that enter the symmetry surfing proposal of \cite{tawe10,tawe11,tawe13}
and that are thus relevant for \cite{tawe12,gakepa16}. In particular, the
properties of the
underlying symmetries of toroidal conformal field theories and their
induced actions on the respective charge lattices are discussed.
Section \ref{nodoom} is devoted to the proof of our claim that
geometric symmetries of toroidal conformal field 
theories, in particular those that
enter symmetry surfing, are not
doomed to fail.
We remark that an alternative, just as immediate proof follows
from the discussion around equation (4.54) of \cite{ta15}.
\section{Geometric symmetries of K3 theories}\label{geosymm}
The Mathieu Moonshine phenomenon, discovered
by T.~Eguchi, H.~Ooguri and Y.~Tachikawa
\cite{eot10}, predicts the existence of a $\Z_2\times\Z_2$ graded
representation
of the ``small'' $N=4$ superconformal algebra 
of \cite{aetal76} at central charge $c=6$,
whose diagonally $\Z_2$-graded character  yields the complex elliptic 
genus of a K3 surface, and 
which simultaneously furnishes a representation of the Mathieu
group $M_{24}$. This should yield the corresponding \textsl{twisted
twining genera} with their strongly restrictive modular properties.
The existence of such an $M_{24}$-module
was proved by T.~Gannon \cite{ga12}. However,
in addition one expects a compatible structure of a 
super vertex operator algebra on this $M_{24}$-module.
The latter  has not been constructed, so far. Neither has
a satisfactory explanation been found for the existence of
an $M_{24}$-module with all this  structure. 

Our quest for an explanation of these phenomena
has led us to propose that the $M_{24}$-module in question
should arise as a subspace of the spaces of states of K3 theories
which is common to all such theories. The representations
constructed in \cite{tawe12,gakepa16} arise precisely as such 
from the spaces of states of $\Z_2$-orbifold conformal field
theories on K3. We suggest that 
the action of $M_{24}$ 
might then be explained by means of certain symmetry groups of
K3 theories on this  space of generic
states, combined from distinct points of the moduli space.
The latter is our proposal of \textsl{symmetry surfing}.

As  explained in \cite{tawe13,we14,we17}, we expect 
such a space of generic states to arise as the 
large volume limit of a topological half-twist of the space of
states of our
K3 theories (denoted $\mathcal X$ in \cite[\S5]{tawe13},
for example). 
According to A.~Kapustin, it thus may be modelled by the 
cohomology of the  chiral de Rham complex of the underlying
K3 surface \cite{ka05}, which according to
\cite[Prop.~3.7 and Def.~4.1]{bo01} indeed carries the structure of a 
super vertex operator algebra, see also \cite{so17}. 
To be compatible with such a large volume
limit, the symmetries that are relevant to Mathieu Moonshine 
must  be \textsl{geometric} in
the sense that they are induced by geometric symmetries of the underlying
K3 surface, independently of the volume. More precisely, we require these
symmetries to fix a geometric interpretation of our theories according to
\cite{asmo94}.

To characterize such symmetries more specifically,
let us recall the
description of the moduli space of K3 theories, following \cite{asmo94,nawe00}.
Indeed, a K3 theory may be specified by data that determine a hyperk\"ahler structure
on a K3 surface $X$, its volume $V\in\R,\, V>0$, and its $B$-field 
$B\in H^2(X,\R)$. Here, the cohomology $H^\ast(X,\R)$ of K3 is equipped with
the scalar product $\langle\cdot,\cdot\rangle$
of signature $(4,20)$ that is induced by the intersection form.
A hyperk\"ahler structure on $X$ may then be uniquely specified by an oriented,
positive definite three-dimensional subspace $\Sigma\subset H^2(X,\R)$.
Denoting by $\upsilon^0\in H^0(X,\Z)$, $\upsilon\in H^4(X,\Z)$ a generating
pair of vectors for the hyperbolic lattice $H^0(X,\Z)\oplus H^4(X,\Z)$ with 
$\langle\upsilon^0,\upsilon\rangle=1$, the K3 theory in question is 
uniquely specified by the positive definite oriented four-dimensional subspace
of $H^\ast(X,\R)$ which is generated by
$$
\left\{\left. \vphantom{{\textstyle V -  {\langle B,B\rangle\over2}}}
\sigma-\langle B,\sigma\rangle \upsilon \right| \sigma\in\Sigma \right\}
\cup
\left\{ \upsilon^0 + B + \left( {\textstyle V -  {\langle B,B\rangle\over2}} \right) \upsilon\right\}.
$$
As is explained, for example, in \cite{ghv12},
the symmetries in question 
in particular induce lattice automorphisms
of $H^\ast(X,\Z)$ which leave the above four-dimensional
subspace of $H^\ast(X,\R)$ invariant, point-wise. To be compatible
with a large volume limit, this property must hold independently of the
value of $V$. 
It follows that the vector $\upsilon$  must be invariant under our lattice
automorphisms. 
Since our large volume limit is not only independent of the value of
$V$ but solely depends on a choice of complex structure,
in all our works
we have been even more restrictive on the symmetries that enter
symmetry surfing. To call a symmetry \textsl{geometric}, 
we require it to
fix the geometric interpretation, i.e.\ we require that the induced
lattice automorphism fixes
both $\upsilon$ and $\upsilon^0$. 
Thus $B$ must also be fixed, see
\cite[footnotes 18, 19]{tawe13}, \cite[\S4]{tawe12}, and
\cite[\S4.1.1]{gtvw14}.

All the symmetries that have been used in symmetry surfing, so far 
\cite{tawe11,tawe12,tawe13,gakepa16}, are induced from 
symmetries of Kummer surfaces that in turn descend from 
symmetries of the underlying complex torus.
In other words, any
such symmetry is given in terms of the geometric interpretation
of our toroidal theory on some torus
$\R^d/\Lambda$, $d=4$, with B-field
$\wt B$. Here, $\wt B$ is given by a real, skew-symmetric $d\times d$ matrix, 
$\Lambda\subset\R^d$ is a lattice of rank $d$ and
by $\Lambda^\ast\subset\R^d$ we denote its dual after identification
of $\R^d$ with $(\R^d)^\ast$ by means of the Euclidean metric $\cdot$,
that is,
$$
\Lambda^\ast 
= \left\{ \mu\in \R^d \mid \mu\cdot\lambda\in\Z\;
\forall\lambda\in\Lambda \right\} .
$$
The corresponding charge lattice then is
\be\label{chargelattice}
\Gamma(\Lambda,\wt B)
= \left\{ {\textstyle{1\over\sqrt2}} (\mu-\wt B\lambda+\lambda;\mu-\wt B\lambda-\lambda)
\mid (\mu,\lambda)\in\Lambda^\ast\oplus\Lambda 
\right\} \subset \R^{d,d},
\ee
where we use the  standard conventions 
as for example in \cite[(1.11)]{nawe00},
 \cite[(A.3)-(A.5)]{gtvw14}, \cite[(3.3)]{hamo17}, and 
 $\R^{d,d}=\R^d\oplus\R^d$ is equipped with the scalar product $\bullet$
 $$
 \forall (p_l;p_r),\, (p_l^\prime;p_r^\prime) \in \Gamma(\Lambda,\wt B)\colon\qquad
 (p_l;p_r)\bullet (p_l^\prime;p_r^\prime)
 = p_l\cdot p_l^\prime - p_r\cdot p_r^\prime
 $$
 of signature $(d,d)$.  
The geometric  symmetries of the toroidal
superconformal field theory that might potentially be affected
by the findings of \cite{hamo17} are  thus given by linear maps $g\in O(d)$ 
with $g\Lambda=\Lambda$ and $g\wt B=\wt Bg$. The induced action of $g$ on 
the charge lattice is
\be\label{geosymmaction}
\forall p = (p_l; p_r)
\in \Gamma(\Lambda,\wt B)\colon\qquad
g(p) := (gp_l;gp_r).
\ee
For concrete examples relevant to symmetry surfing, the reader
is referred to \cite[\S1]{tawe13}. We emphasize that the respective 
symmetry groups may be non-abelian, and that this does not impose
any additional difficulties.
\section{Not Doomed to Fail}\label{nodoom}
Consider a toroidal conformal field theory with charge lattice
$\Gamma\subset\R^{d,d}$, and a lattice automorphism
$\gamma$ of $\Gamma$ which fixes the parameter point of the
theory, i.e.\ which acts  on the charge lattice by means of
$$
\forall p = (p_l; p_r)
\in \Gamma\colon\qquad
\gamma(p) = (g_lp_l;g_rp_r)
$$
with $g_l,\,g_r\in O(d)$. Assume 
that $\gamma$ has order $\ell$. Then, reevaluating 
obstructions previously discussed in different interpretations
or contexts \cite{nsv87,nsv91,le85,bhl06,ta15},
J.~Harvey and G.~Moore find the following obstruction  
for $\gamma$ to lift to an automorphism of order $\ell$ of the
corresponding toroidal conformal field theory
that leaves invariant winding-momentum fields associated to 
$\gamma$-invariant charge vectors \cite[(2.17)]{hamo17}:
such a lift is doomed to fail
if $\ell$ is even and
\be\label{doom}
\exists p\in\Gamma\colon\qquad
p\bullet \gamma^{\ell/2}(p)\notin2\Z.
\ee
To show that the geometric symmetries of toroidal conformal
field theories  are not doomed to fail,
we may thus assume without loss of generality that $\ell=2$.
As explained in Section \ref{geosymm} above, 
by (\ref{geosymmaction}) we furthermore assume  
that $g_l=g_r=g\in O(d)$, $\Gamma=\Gamma(\Lambda,\wt B)$ as in (\ref{chargelattice})
with $\wt B^T=-\wt B$, and that $g\Lambda=\Lambda$,
$g\wt B=\wt Bg$. We thus have $g=g^{-1}=g^T$, and we find
$$
(g\wt B)^T = -\wt Bg=-g\wt B.
$$ 
In particular, we have
\be\label{gBisantisymm}
\forall \lambda,\,\lambda^\prime\in\R^d\colon\qquad
\lambda\cdot (g\wt B)\lambda^\prime + \lambda^\prime\cdot (g\wt B)\lambda =0.
\ee
For charge vectors 
\be\label{choices}
\textstyle
p= {1\over\sqrt2} (\mu-\wt B\lambda+\lambda,\mu-\wt B\lambda-\lambda),\quad
p^\prime= {1\over\sqrt2} (\mu^\prime-\wt B\lambda^\prime+\lambda^\prime,\mu^\prime-\wt B\lambda^\prime-\lambda^\prime)
\ee
with arbitrary $\lambda,\,\lambda^\prime\in\Lambda$ and $\mu,\,\mu^\prime\in\Lambda^\ast$
we thus have
\begin{eqnarray*}
p^\prime\bullet \gamma(p) 
&=& (\mu^\prime-\wt B\lambda^\prime)\cdot g\lambda + \lambda^\prime\cdot g(\mu-\wt B\lambda)\\
&\stackrel{g=g^T}{=}& 
\mu^\prime\cdot g\lambda + \mu\cdot g\lambda^\prime
- \lambda\cdot (g\wt B)\lambda^\prime - \lambda^\prime\cdot (g\wt B)\lambda\\
&\stackrel{(\footnotesize\ref{gBisantisymm})}{=}&
\mu^\prime\cdot g\lambda + \mu\cdot g\lambda^\prime.
\end{eqnarray*}
In particular, 
$$
\forall 
\textstyle
p= {1\over\sqrt2} (\mu-\wt B\lambda+\lambda,\mu-\wt B\lambda-\lambda)
\in\Gamma(\Lambda,\wt B)\colon\qquad
p\bullet \gamma(p)  = 2 \mu\cdot g\lambda \in2\Z
$$
since by assumption, $g\lambda\in\Lambda$ and $\mu\in\Lambda^\ast$.
In other words, the ``doomed to fail'' condition (\ref{doom}) does not hold.
In particular, at $d=4$ we learn that the symmetries that have been 
relevant for symmetry surfing, so far, are \textsl{not} doomed to fail.

We remark that the ``doomed
to fail condition'' of \cite{hamo17} is solely testing the 
\textsl{cyclic subgroups} of a given symmetry group. However,
as noted  at the end of Section \ref{geosymm}, symmetry-surfing
does involve non-cyclic, in fact even non-abelian symmetry groups.
Actually, all
examples of symmetries that \textsl{are} doomed to fail and that are discussed in 
\cite{hamo17} arise in non-abelian global symmetry groups of special conformal 
field theories with enhanced 
symmetry.
This raises the question of
whether every geometric symmetry group $G$ has a lift to a symmetry group
of the respective conformal field theory which is isomorphic to $G$.
That this is indeed the case follows from the existence of an invariant $2$-cocycle
$$
\varepsilon\colon \Gamma\times\Gamma\longrightarrow\left\{\pm1\right\}
$$
on the charge lattice $\Gamma$ which governs the operator product
expansions between vertex operators (see, for example, \cite{frka80,se81,gool84,ka96} for
the classical results). Indeed, with notations as in \eqref{choices}, one may use
the cocycle
$$
\forall p,\,p^\prime\in\Gamma\colon\qquad 
\varepsilon\!\left(p,p^\prime\right) := (-1)^{\mu\cdot\lambda^\prime}.
$$
For a geometric symmetry $\gamma$ as above, $g\in O(d)$ thus implies 
$$
\forall p,\,p^\prime\in\Gamma\colon\qquad 
\varepsilon\!\left(\gamma(p),\gamma(p^\prime)\right) = \varepsilon\!\left(p,p^\prime\right).
$$
In terms of \cite[Appendix A]{hamo17}, this means that a lift $G\longrightarrow\widehat G$,
$g\mapsto T_g$, of our symmetry group $G$ exists which obeys
$T_{g_1}\circ T_{g_2} = T_{g_1g_2}$ for all $g_1,\, g_2\in G$
(see \cite[(A.5)--(A.11)]{hamo17}), thus yielding $G\cong\widehat G$.
\vskip 1cm

{\textbf{Acknowledgements}

We thank  J.~Harvey and G.~Moore  for communicating
a preliminary version of their  work \cite{hamo17} to us very early 
on, and for very useful discussions. This gave us the chance to set the record straight  quickly,
regarding symmetry surfing not being doomed to fail.
We also thank an anonymous referee for carefully reading the
manuscript and for raising a number of questions whose answers will
certainly help the readers to put our work in the right context.
%
%
%
%
%
%
%
\def\polhk#1{\setbox0=\hbox{#1}{\ooalign{\hidewidth
  \lower1.5ex\hbox{`}\hidewidth\crcr\unhbox0}}} \def\cprime{$^\prime$}
  \newcommand{\noopsort}[1]{}

\providecommand{\href}[2]{#2}\begingroup\raggedright\endgroup
\end{document}